\documentclass[12pt,a4paper]{article}

\usepackage{amsmath}
\usepackage{amsfonts}
\usepackage{amssymb}
\usepackage{amsthm}
\usepackage{latexsym}

\makeatletter
\@addtoreset{equation}{section}
\makeatother

\newcommand{\be}{\begin{equation}}
\newcommand{\ee}{\end{equation}}
\newcommand{\nnr}{\nonumber \\}

\newcommand{\eq}[1]{(\ref{#1})}
\newcommand{\fr}{\frac}
\newcommand{\tf}{\tfrac}

\newcommand{\df}{\textrm{d}}
\newcommand{\expe}[1]{\textrm{e}^{#1}}
\newcommand{\pd}{\partial}
\newcommand{\sr}{\sqrt}

\begin{document}


\thispagestyle{empty}

\vspace*{100pt}
\begin{center}
\textbf{\Large{Characterization of three-dimensional Lorentzian metrics that admit four Killing vectors}}\\
\vspace{50pt}
\large{David D. K. Chow}
\end{center}

\begin{center}
{\bf Abstract\\}
\end{center}
We consider three-dimensional Lorentzian metrics that locally admit four independent Killing vectors.  Their classification is summarized, and conditions for characterizing them are found.  These consist of algebraic classification of the traceless Ricci tensor, and other conditions satisfied by the curvature and its derivative.
\newpage


\section{Introduction}


The Lorentzian signature metrics in three spacetime dimensions that locally admit four independent Killing vectors have been classified long ago by Kruchkovich \cite{kruchk, kruchk2}.  These metrics are not only of mathematical interest, but several have appeared in studies of 3-dimensional gravity, which can be a toy model for insights into quantum gravity, and as building blocks for metrics in higher dimensions.  The issue of characterizing the spacetimes with a 4-dimensional isometry group was particularly considered by Bona and Coll \cite{boncol}, and some other results are scattered across the mathematics and physics literature.  More recently, the topic has been revisited, in the context of a general algorithm for counting the number of Killing vectors in a 3-dimensional spacetime, by Nozawa and Tomoda \cite{Nozawa:2019dwu}.  In this article, we provide a summary of the classification, obtain some further results to characterize the different classes, and provide some illustrative examples.

In pure three-dimensional Einstein gravity, which lacks a propagating graviton, all solutions are locally maximally symmetric, with six Killing vectors locally, and so any interesting features must come from global identifications.  A richer class of three-dimensional gravitational theories can come from coupling to matter or by considering higher-derivative gravitational terms.  In these more general theories, there are wider classes of admissible local solutions.  It is not possible to realize exactly five Killing vectors locally, and so four Killing vectors is the next to be realized.  

In three spacetime dimensions one often considers global identifications that alter the number of globally defined Killing vectors.  For example, the BTZ black hole \cite{Banados:1992wn} is locally anti-de Sitter (AdS) spacetime, but the number of Killing vectors is reduced from 6 to 2 \cite{Banados:1992gq}, the same as for the 4-dimensional Kerr--AdS black hole.  The self-dual Coussaert--Henneaux solution, which does not represent a black hole, is an example in which the 6 Killing vectors of anti-de Sitter spacetime are reduced to 4 global Killing vectors upon global identification \cite{Coussaert:1994tu}.  In contrast, in this article we are only concerned with the local geometry.

A useful result for practical classification of the metrics with 4 Killing vectors is due to Bona and Coll \cite{boncol}.  It states that a 3-dimensional Lorentzian metric has a 4-dimensional isometry group if and only if the Ricci tensor takes the form
\be
R_{a b} = \lambda_1 g_{a b} + \lambda_2 u_a u_b ,
\ee
where $\lambda_1$ and $\lambda_2 \neq 0$ are constants, and either: $u_a$ is a Killing vector, i.e.\ $\nabla_{(a} u_{b)} = 0$, that is non-null (which can be normalized so that $u^a u_a = \pm 1$) or null; or $u_a$ is a recurrent null vector, i.e.\ $\nabla_a u_b = \alpha u_a u_b$, $u^a u_a = 0$, for some constant $\alpha$.  The traceless Ricci tensor $S_{a b} = R_{a b} - \tf{1}{3} R g_{a b}$ in the timelike, spacelike and null cases then takes the respective forms
\begin{align}
& \textrm{Type D$_t$:}& S_{a b} & = \lambda (g_{a b} + 3 t_a t_b) , & t^a t_a & = - 1 , \nnr
& \textrm{Type D$_s$:}& S_{a b} & = \lambda (g_{a b} - 3 s_a s_b) , & s^a s_a & = 1 , \nnr
& \textrm{Type N:}& S_{a b} & = \lambda k_a k_b , & k^a k_a & = 0 ,
\label{DDN}
\end{align}
for some constant $\lambda$.

In the case of a Killing vector, either null or non-null, the assumptions of Bona and Coll imply that the Ricci tensor is cyclic-parallel \cite{gray},
\be
\nabla_{(a} R_{b c)} = 0 .
\ee
In the case of a recurrent null vector, the assumptions imply that the Ricci tensor satisfies the Codazzi condition \cite{gray}
\be
\nabla_{[a} R_{b] c} = 0 .
\ee
Recall that a 3-dimensional metric is conformally flat if and only if its Cotton tensor $C_{a b}$ vanishes, where
\be
C_{a b} = \epsilon{_a}{^{c d}} \nabla_c (R_{d b} - \tf{1}{4} R g_{b d}) ,
\ee
which is a symmetric, traceless and divergence-free (pseudo)tensor.  The vanishing of $C_{a b}$ is equivalent to
\be
\nabla_{[a} R_{b] c} - \tf{1}{4} \nabla_{[a} R g_{b] c} = 0 ,
\ee
and so, if the Ricci scalar is constant, the Codazzi condition is equivalent to conformal flatness.  In both null and non-null cases, if $u_a$ is parallel, i.e.\ covariantly constant with $\nabla_a u_b = 0$, then the spacetime is symmetric,
\be
\nabla_a R_{b c} = 0 .
\ee

Whereas the Bona and Coll assumptions imply consequences such as the cyclic-parallel condition and conformal flatness, in contrast, we will obtain in this article some results in which these consequences are instead taken to be assumptions.  One further consequence of the Bona and Coll assumptions in the case of a recurrent null vector with $S_{a b} = \lambda k_a k_b$, $\nabla_a k_b = \alpha k_a k_b$, is the relation $\lambda \, \nabla_a S_{b c} \, \nabla_d S_{e f} = \alpha^2 \, S_{a b} S_{c d} S_{e f}$, and we shall again provide a converse result.

As indicated in \eq{DDN}, if the traceless Ricci tensor can be decomposed as shown, but more generally allowing $\lambda$ to be a function of position, the traceless Ricci tensor is said to have Segre type D$_t$, D$_s$ and N respectively.  This fits into a wider classification of rank-2 tensors $S{^a}{_b}$, with further types III, II and I that we do not require here.  Type O is defined by $S_{a b} = 0$, which in 3 dimensions is equivalent to the metric being Einstein.  For further details, see e.g.\ \cite{Chow:2009km}.  A metric with Segre type D$_\textrm{t}$ could be interpreted as a solution of Einstein gravity coupled to matter that is a certain perfect fluid, while type N could be interpreted as null radiation.  We have presented the classification in \eq{DDN} in terms of the existence of a certain orthonormal or null triad for decomposing $S_{a b}$, but the classification is also equivalent to a Jordan normal form classification of $S{^a}{_b}$ regarded as an endomorphism.  Therefore, the classification can also be expressed in terms of the minimal polynomial for $S{^a}{_b}$.  Type N is equivalent to $S{^a}{_c} S{^c}{_b} = 0$, and types D$_\textrm{t}$ and D$_\textrm{s}$ are equivalent to $S{^a}{_c} S{^c}{_b} + \lambda S{^a}{_b} - 2 \lambda^2 \delta^a_b = 0$, which are straightforward to check for a given metric.

One situation in which spacetimes with 4 Killing vectors have appeared is in topologically massive gravity (TMG) \cite{Deser:1982vya, Deser:1982vyb, Deser:1982sv}.  This is a 3-dimensional theory of gravity whose field equations consist of Einstein gravity, i.e.\ the Einstein tensor $G_{a b} = R_{a b} - \tf{1}{2} R g_{a b}$ and a possible cosmological constant term $\Lambda g_{a b}$, coupled to the Cotton tensor:
\be
G_{a b} + \Lambda g_{a b} + \fr{1}{\mu} C_{a b} = 0 .
\label{TMG0}
\ee
Note that the Ricci scalar $R = 6 \Lambda$ is constant, and that $\mu$ changes sign under a change of orientation.  As usual for the physics literature, there are two dimensionful constants, $\Lambda$ and $\mu \neq 0$, which are equivalent to a single dimensionless constant.  Equivalently, in terms of the traceless Ricci tensor, we have
\be
S_{a b} + \fr{1}{\mu} C_{a b} = 0 ,
\label{TMG}
\ee
where we may now regard the single constant $\mu$ as dimensionless.  Using the contracted Bianchi identity $\nabla_a G^{a b} = 0$, \eq{TMG} implies that the Ricci scalar $R$ is constant.  The theory has seen interest as a toy model of gravity in 3 dimensions, with the advantage over pure Einstein gravity of having a propagating graviton resuting from the higher-derivative Cotton tensor term.  However, for the purposes of this article, we use the TMG equation only as a mathematical statement of the vanishing of some linear combination (with constant coefficients) of the Einstein tensor, the metric, and the Cotton tensor.  It turns out that all of the 3-dimensional spacetimes that admit 4 Killing vectors either solve the equations of TMG, or are conformally flat, which is effectively the $\mu = 0$ limit of TMG.

Algebraic conditions on the curvature have been used practically in the study of TMG solutions.  It has been shown that many such solutions in the literature are locally equivalent to a small set \cite{Chow:2009km} (hundreds of references are given in the remarkable book \cite{Garcia-Diaz:2017cpv}).  Their multiple rediscoveries is likely a consequence of their high symmetry.

In particular, it was previously shown \cite{Chow:2009km} that all type D solutions of TMG correspond to squashed AdS, or certain generalizations.  Some of the spacetimes with 4 Killing vectors are symmetric, and these have also been classified \cite{calvaruso}.  We further generalize, providing characterizations of all cases in which there are 4 Killing vectors.

Unlike the result of Bona and Coll, we do not assume the existence of a Killing vector or recurrent null vector, or that the eigenvalues of $S{^a}{_b}$ are constant.  We instead make assumptions on the algebraic type of $S{^a}{_b}$ and on equations directly satisfied by the curvature, i.e.\ without using a decomposition into a triad.

The outline of this paper is as follows.  In Section 2, we summarize the various classes of Lorentzian metrics that admit 4 Killing vectors.  We write the metrics and Killing vectors in a canonical form.  In Section 3, we provide the classification results, along the way finding the general metrics that are type N and conformally flat.  In Section 4, we illustrate how these results can be used to provide explicit matching between some of the original metrics of Kruchkovich and the canonical forms provided here.


\section{Metrics with four Killing vectors}


The spacetimes with 4 Killing vectors can be classified in particular according to: their Segre type, i.e.\ the algebraic classification of the traceless Ricci tensor; whether they are conformally flat or solve TMG; and whether the Ricci tensor is cyclic-parallel.  We summarize some properties in Table \ref{summarytable}.  There are perhaps three natural broad classes of metrics.
\begin{table}
\centering
\begin{tabular}{|c|c|c|c|}
\hline
Metric & Segre type & CF or TMG & $\nabla_{(a} R_{b c)} = 0$? \\
\hline
Timelike-squashed AdS & D$_\textrm{t}$ & TMG & Yes \\
Spacelike-squashed AdS & D$_\textrm{s}$ & TMG & Yes \\
$\mathbb{R} \times S^2$, $\mathbb{R} \times H^2$ & D$_\textrm{t}$ & CF & Yes \\
$\textrm{(A)dS}_2 \times \mathbb{R}$ & D$_\textrm{s}$ & CF & Yes \\
\hline
Null warped AdS & N & TMG & Yes \\
\hline
Symmetric plane wave & N & CF & Yes \\
Non-symmetric plane wave & N & CF & No \\
\hline
\end{tabular}
\caption{Spacetimes with 4 Killing vectors; ``CF'' means conformally flat, ``TMG'' means equations of TMG satisfied}
\label{summarytable}
\end{table}

The first class consists of metrics known as squashed (or warped) anti-de Sitter, and slight generalizations.  The timelike-squashed metrics are timelike fibrations over a 2-dimensional constant curvature Euclidean-signature space, such as $H^2$, whereas the spacelike-squashed metrics are spacelike fibrations over a 2-dimensional constant curvature Lorenztian-signature spacetime, such as AdS$_2$.  Squashed AdS can be regarded as an analytic continuation of the Berger sphere.  The Segre types of timelike-and spacelike-squashed AdS are respectively D$_\textrm{t}$, D$_\textrm{s}$, and their isometry algebras are semisimple.

The second class is null warped AdS, which is a pp-wave generalization of AdS.  It has Segre type N, solves TMG for a special coupling, has a cyclic-parallel Ricci tensor, and has a semisimple isometry algebra.

The third class consists of special plane wave metrics, and in particular possess a recurrent null vector.  They are divided into those that are symmetric, i.e.\ $\nabla_a R_{b c} = 0$, and those that are not.  They are conformally flat, are of Segre type N, and their isometry groups are solvable.

We also summarize the Killing vectors, presenting them to match canonical forms of the corresponding Lie algebras as given in \cite{Patera:1976ud} (real 4-dimensional Lie algebras are also reviewed in \cite{maccal}).  They have also been discussed in \cite{Nozawa:2019dwu}, however we have used coordinate freedom to fix the metric as far as possible.  These may be useful for detailed comparison between metrics written in different coordinate systems.


\subsection{Type D}



\subsubsection{Conformally flat: $\mathbb{R} \times \Sigma_2$}


These are direct products of a 2-dimensional maximally symmetric space(time) $\Sigma_2$ and $\mathbb{R}$.  Either $\Sigma_2$ is Euclidean signature and the $\mathbb{R}$ is timelike, or $\Sigma_2$ is Lorentzian signature and the $\mathbb{R}$ is spacelike.  Each case has 3 possibilities, depending on whether the curvature of $\Sigma_2$ is positive, negative or zero, making a total of 6 distinct cases.  However, if $\Sigma_2$ is flat, then we have Minkowski spacetime, which is type O rather than type D, which we exclude here, but will consider again when we consider a one-parameter generalization.  The maximally symmetric $\Sigma_2$ gives 3 Killing vectors, and the $\mathbb{R}$ gives 1 more Killing vector.  The spacetime is symmetric, i.e.\ $\nabla_a R_{b c} = 0$, and so the metric is conformally flat.

As an explicit example, consider a Euclidean signature $\Sigma_2$, taken to be hyperbolic space $H^2$, and a timelike $\mathbb{R}$.  The metric can therefore be written locally as
\be
\df s^2 = - \df t^2 + \df \theta^2 + \sinh^2 \theta \, \df \phi^2 .
\ee
The Ricci scalar is $R = -2$.  The traceless Ricci tensor, which is type D$_\textrm{t}$, is
\be
S_{a b} = - \tf{1}{3} (g_{a b} + 3 t_a t_b) ,
\ee
where $t^a \, \pd_a = \pd_t$ is a unit timelike vector that is covariantly constant, i.e.\ $\nabla_a t_b = 0$.


\subsubsection{Not conformally flat: squashed metrics}


These metrics generalize each of the six $\mathbb{R} \times \Sigma_2$ cases to include one extra dimensionless parameter.  For timelike-squashings, these metrics take the form
\be
\df s^2 = - (\df t + \mathcal{A})^2 + \df s_2^2 ,
\label{timelikesquash}
\ee
and for spacelike-squashings, these metrics take the form
\be
\df s^2 = \df s_2^2 + (\df z + \mathcal{A})^2 ,
\label{spacelikesquash}
\ee
where $\df \mathcal{A} = \tf{2}{3} \mu \epsilon_2$, with $\mu \neq 0$, and $\epsilon_2$ is the volume-form of $\df s_2^2$, which is a constant curvature space(time) with canonical normalization.  The limit $\mu = 0$ gives the $\mathbb{R} \times \Sigma_2$ metrics.  Again, $\Sigma_2$ gives 3 Killing vectors, and the $\mathbb{R}$ gives another.  The timelike-squashed metrics are Segre type D$_\textrm{t}$, and the spacelike-squashed metrics are type D$_\textrm{s}$.  The Ricci scalar is constant, and the Ricci tensor is cyclic-parallel, i.e.\ $\nabla_{(a} R_{b c)} = 0$.  The metric satisfies the TMG equation \eq{TMG}.

The metrics with zero or positive curvature $\Sigma_2$ do not contain any Einstein limit, and have strictly positive Ricci scalars.  However, for want of a better terminology, they have been dubbed ``warped flat'' and ``warped de Sittter'' \cite{Anninos:2009jt}.

As an explicit example, we consider the one-parameter generalization of $\mathbb{R} \times \mathbb{H}^2$.  This can be written locally as
\be
\df s^2 = - ( \df t + \tf{2}{3} \mu \,  \cosh \theta \, \df \phi ) ^2 + \df \theta^2 + \sinh^2 \theta \, \df \phi^2 .
\ee
The Ricci scalar is $R = - 2 + \tf{2}{9} \mu^2$, and the traceless Ricci tensor is
\be
S_{a b} = (\tf{1}{9} \mu^2 - \tf{1}{3}) (g_{a b} + 3 t_a t_b) ,
\ee
where $t^a \, \partial_a = \partial_t$ is a unit timelike vector.  Since $t^a$ satisfies $\nabla_a t_b = \tf{1}{3} \mu \epsilon_{a b c} t^c$, it is a Killing vector. 

These metrics have been considered in more detail elsewhere; some examples are \cite{Rooman:1998xf, dulupo, Bengtsson:2005zj, Anninos:2008fx}.  Within the TMG literature, the solution was rediscovered many times in a range of coordinate systems, as illustrated in \cite{Chow:2009km} (reused in Chapter 17 of \cite{Garcia-Diaz:2017cpv}).  The timelike-squashed AdS solution has also been considered as a solution of Einstein gravity coupled to matter; a recent example considers pressureless dust \cite{Remmen:2018oqw}.


\subsection{Type N, not conformally flat: null warped AdS}


The metric can be expressed as
\be
\df s^2 = \df \rho^2 + 2 \expe{2 \rho} \, \df u \, \df v \pm \expe{4 \rho} \, \df u^2 .
\label{nullwarpedAdS}
\ee
The Ricci scalar is $R = - 6$, and the traceless Ricci tensor and Cotton tensor are
\begin{align}
S_{a b} & = \mp 4 k_a k_b , & C_{a b} & = \mp 12 k_a k_b ,
\end{align}
where $k^a \, \partial_a = \partial_v$ is a null Killing vector that is not covariantly constant.  We have chosen the orientation $\epsilon_{v u \rho} = 1$.  It follows that the TMG equation with a special coupling is satisfied,
\be
S_{a b} - \tf{1}{3} C_{a b} = 0 ,
\ee
and the Ricci tensor is cyclic-parallel, $\nabla_{(a} R_{b c)} = 0$.  However, the spacetime is not symmetric, $\nabla_a R_{b c} \neq 0$.  There are four independent Killing vectors:
\begin{align}
K_1 & = \partial_v , & K_2 & = 2 u \, \partial_u - \partial_\rho , & K_3 & = \tf{1}{2} \expe{- 2  \rho} \, \partial_v - u^2 \, \partial_u + u \, \partial_\rho , & K_4 & = \partial_u .
\end{align}
Their non-trivial commutators are
\begin{align}
[K_2, K_3] & = 2 K_3 , & [K_2, K_4] & = - 2 K_4 , & [K_3, K_4] & = K_2 ,
\end{align}
which gives the isometry group $\textrm{SL} (2, \mathbb{R}) \times \textrm{U}(1)$.  The metric is known as null warped AdS and has appeared in a number of contexts.  For example, it has been considered as part of a string background \cite{Israel:2004vv, Detournay:2005fz}, within geometries that realize non-relativistic symmetries \cite{Son:2008ye, Balasubramanian:2008dm}, and in the context of TMG \cite{Anninos:2008fx, Anninos:2010pm}.


\subsection{Type N, conformally flat: special plane waves}


The other classes of metrics admitting 4 Killing vectors are certain plane waves.  In 3 dimensions, plane waves generically admit 3 Killing vectors \cite{Torre:1999ye}.  However, there are two classes of special cases in which there is an additional Killing vector.  These are furthermore examples of Walker manifolds \cite{walker}, which admit a recurrent null vector, i.e.\ $\nabla_a k_b = \alpha k_a k_b$ for some null $k_a$ and constant $\alpha$.  This is precisely the null vector in the decomposition of the type N traceless Ricci tensor $S_{a b} = \pm k_a k_b$.  The distinction between the two classes is whether $\alpha = 0$ or $\alpha \neq 0$.


\subsubsection{Symmetric Walker metric}


The metric can be written as
\be
\df s^2 = \df \rho^2 + 2 \, \df u \, \df v \pm \rho^2 \, \df u^2 .
\label{symmetricWalker}
\ee
Note that $\partial_u$ gives an additional Killing vector compared to a generic plane wave.  The Ricci scalar vanishes, $R = 0$, and the traceless Ricci tensor is type N, with
\be
S_{a b} = \mp k_a k_b ,
\ee
where $k^a \partial_a = \partial_v$ is a null vector that is covariantly constant, $\nabla_a k_b = 0$, which defines a pp-wave.  This implies that the spacetime is symmetric, $\nabla_a R_{b c} = 0$, and so conformally flat.  Theorem 6 of \cite{chgava} classifies the symmetric Walker metrics, presented with two additional functions of one variable, which can be reduced to the canonical form \eq{symmetricWalker} by coordinate transformations of the form \eq{transform}.

With the positive sign in \eq{symmetricWalker}, the Killing vectors
\begin{align}
K_1 & = \partial_v , & K_2 & = \tf{1}{\sr{2}} \expe{-u} (\rho \, \partial_v + \partial_\rho) , & K_3 & = \tf{1}{\sr{2}}  \expe{u} (\rho \, \partial_v - \partial_\rho) , & K_4 & = \partial_u
\end{align}
have the non-trivial commutators
\begin{align}
[K_2 , K_3] & = K_1 , & [K_2 , K_4] & = K_2 , & [K_3 , K_4] & = - K_3 ,
\end{align}
which is the solvable algebra $A_{4, 8}$ in the notation of \cite{Patera:1976ud}.  Let $L_i$ be the 1-forms corresponding to the Killing vectors $K_i$, so
\begin{align}
L_1 & = \df u , & L_2 & = \tf{1}{\sr{2}} \expe{-u} (\rho \, \df u + \df \rho) , & L_3 & = \tf{1}{\sr{2}} \expe{u} (\rho \, \df u - \df \rho) , & L_4 & = \df v + \rho^2 \, \df u .
\end{align}
The metric is
\be
\df s^2 = 2 (L_1 L_4 - L_2 L_3) = 2 \, \df u \, (\df v + \df \mathcal{A}) + \df s_2^2 ,
\ee
where $\df s_2^2 = \df \rho^2 - \rho^2 \, \df u^2$ is Minkowski$_2$ spacetime with volume-form $\epsilon_2 = \tf{1}{2} \df \mathcal{A} = \rho \, \df \rho \wedge \df u$.

With the negative sign in \eq{symmetricWalker}, the Killing vectors
\begin{align}
K_1 & = \partial_v , \qquad K_2 = \sin u \, \partial_\rho - \rho \, \cos u \, \partial_v , & K_3 & = \cos u \, \partial_\rho + \rho \, \sin u \, \partial_v , \qquad K_4 = \partial_u
\end{align}
have the non-trivial commutators
\begin{align}
[K_2 , K_3] & = K_1 , & [K_2 , K_4] & = - K_3 , & [K_3 , K_4] & = K_2 ,
\end{align}
which is the solvable algebra $A_{4, 10}$ in the notation of \cite{Patera:1976ud}.  The corresponding 1-forms are
\begin{align}
L_1 & = \df u , & L_2 & = \sin u \, \df \rho - \rho \, \cos u \, \df u , & L_3 & = \cos u \, \df \rho + \rho \, \sin u \, \df u , & L_4 & = \df v - \rho^2 \, \df u .
\end{align}
The metric is
\be
\df s^2 = 2 L_1 L_4 + L_2^2 + L_3^2 = 2 \, \df u \, (\df v + \mathcal{A}) + \df s_2^2 ,
\ee
where $\df s_2^2 = \df \rho^2 + \rho^2 \, \df u^2$ is $\mathbb{R}^2$ with volume-form $\epsilon_2 = \tf{1}{2} \df \mathcal{A} = \rho \, \df \rho \wedge \df u$.


\subsubsection{Non-symmetric Walker metric}


These metrics can be written as
\be
\df s^2 = \df \rho^2 + 2 \, \df u \, \df v + \fr{c \rho^2}{u^2} \, \df u^2 ,
\label{nonsymmetricWalker}
\ee
for some constant $c \neq 0$.  Note that there is an additional scaling symmetry $u \rightarrow \lambda u$, $v \rightarrow \lambda^{-1} v$, which, compared to a generic plane wave, gives an additional Killing vector, $u \, \partial_u - v \,\partial_v$.  The Ricci scalar is $R = 0$.  The traceless Ricci tensor is
\be
S_{a b} = - c k_a k_b ,
\ee
where $k^a \, \partial_a = u^{-1} \, \partial_v$ is a recurrent null vector that satisfies
\be
\nabla_a k_b = - k_a k_b .
\ee
It follows that the Ricci tensor satisfies the Codazzi condition $\nabla_a R_{b c} = \nabla_b R_{a c}$, which implies, using the contracted Bianchi identity, the conformal flatness condition $C_{a b} = 0$.  However, the Ricci tensor is not cyclic-parallel, $\nabla_{(a} R_{b c)} \neq 0$, and so the spacetime is not symmetric, $\nabla_a R_{b c} \neq 0$.

Two Killing vectors take a universal form, independent of the value of $c$:
\begin{align}
K_1 & = \partial_v , & K_4 & = u \partial_u - v \partial_v .
\end{align}
The other two Killing vectors are of the form
\begin{align}
K_2 & = S_2 (u) \, \partial_\rho - S_2'(u) \, \partial_v , & K_3 & = S_3 (u) \, \partial_\rho - S_3'(u) \, \partial_v , 
\label{K2K3}
\end{align}
where $S_2$ and $S_3$ are independent solutions of $S''(u) - c u^{-2} S(u) = 0$ \cite{Torre:1999ye}.

For $c = \tf{1}{4} (k^2 - 1) > - \tf{1}{4}$, with $k > 0$, the remaining Killing vectors are
\begin{align}
K_2 & = u^{(k + 1)/2} \partial_\rho - \tf{1}{2} (k + 1) \rho u^{(k - 1)/2} \partial_v , & K_3 & = u^{(1 - k)/2} \partial_\rho + \tf{1}{2} (k - 1) \rho u^{- (k + 1)/2} \partial_v .
\end{align}
These obey the commutation relations
\begin{align}
[K_2, K_3] & = k K_1 , & [K_1, K_4] & = - K_1 , & [K_2, K_4] & = - \tf{1}{2} (k + 1) K_2 , & [K_3, K_4] & = \tf{1}{2} (k - 1) K_3 . 
\end{align}
If we set $e_1 = - k K_1$, $e_2 = - K_2$, $e_3 = K_3$, $e_4 = - \fr{2}{k + 1} K_4$, and $b = (1 - k)/(1 + k) \in (-1, 1]$, we see that this is the algebra $A^b_{4, 9}$ in the notation of \cite{Patera:1976ud}.

For $c = - \tf{1}{4}$, the remaining Killing vectors are
\begin{align}
K_2 & = u^{1/2} \, \partial_\rho - \tf{1}{2} \rho u^{-1/2} \, \partial_v , & K_3 & = u^{1/2} \log u \, \partial_\rho - \rho u^{-1/2} (1 + \tf{1}{2} \log u) \, \partial_v .
\end{align}
These obey the commutation relations
\begin{align}
[K_2, K_3] & = - K_1 , & [K_1, K_4] & = K_1 , & [K_2, K_4] & = \tf{1}{2} K_2 , & [K_3, K_4] & = K_2 + \tf{1}{2} K_3 .
\end{align}
If we set $e_1 = - 2 K_1$, $e_2 = - 2 K_2$, $e_3 = - K_3$, $e_4 = 2 K_4$, we see that this is the algebra $A_{4, 7}$ in the notation of \cite{Patera:1976ud}.

For $c = - \tf{1}{4} (k^2 + 1) < - \tf{1}{4}$ with $k > 0$, the remaining Killing vectors are \eq{K2K3} with $S_2 = \sr{u} \cos (\tf{1}{2} k \log u)$, $S_3 = \sr{u} \sin (\tf{1}{2} k \log u)$.
The commutation relations are
\begin{align}
[K_2, K_3] & = - \tf{1}{2} k K_1 , & [K_1, K_4] & = - K_1 , \nonumber \\
[K_2, K_4] & = - \tf{1}{2} K_2 + \tf{1}{2} k K_3 , & [K_3, K_4] & = - \tf{1}{2} K_3 - \tf{1}{2} k K_2 .
\end{align}
If we set $e_1 = - 2 K_1$, $e_2 = \fr{2}{\sr{k}} K_2$, $e_3 = \fr{2}{\sr{k}} K_3$, $e_4 = - \fr{2}{k} K_4$, and $a = 1/k > 0$, we see that this is the algebra $A^a_{4, 11}$ in the notation of \cite{Patera:1976ud}.


\section{From algebraic type to four Killing vectors}


The spacetimes admitting 4 Killing vectors can be characterized in terms of the algebraic type of the traceless Ricci tensor and other algebraic properties of the curvature and its derivative as follows:
\begin{enumerate}
\item A 3-dimensional conformally flat, symmetric spacetime with Segre type D$_\textrm{t}$ is $\mathbb{R} \times S^2$ or $\mathbb{R} \times H^2$, and with type D$_\textrm{s}$ is $\textrm{AdS}_2 \times \mathbb{R}$ or $\textrm{dS}_2 \times \mathbb{R}$.
\item A solution of TMG with Segre type D$_\textrm{t}$ is a timelike-squashed metric \eq{timelikesquash}, and with Segre type D$_\textrm{s}$ is a spacelike-squashed metric \eq{spacelikesquash}.
\item A solution of TMG with Segre type N and a cyclic-parallel Ricci tensor is null warped AdS \eq{nullwarpedAdS}.
\item A 3-dimensional conformally flat, symmetric spacetime with Segre type N is the symmetric conformally flat Walker metric \eq{symmetricWalker}.
\item A 3-dimensional conformally flat spacetime with Segre type N that satisfies, for some constant $c \neq 0$, $\nabla_a R_{b c} \, \nabla_d R_{e f} + 4 c^{-1} S_{a b} S_{c d} S_{e f} = 0$ is the non-symmetric conformally flat Walker metric \eq{nonsymmetricWalker}.
\end{enumerate}
1, 2 and 4 are standard results in the literature, which we summarize first.

The type D result for TMG was proved in \cite{Chow:2009km}.  The first part of the proof shows that the eigenvalues of the Ricci tensor are constant.  The second part of the proof uses the TMG equations, but one could alternatively use the result of Bona and Coll \cite{boncol}.

The symmetric condition $\nabla_a R_{b c} = 0$ is a strong constraint.  3-dimensional symmetric spacetimes were classified in \cite{calvaruso}, and can be split into 3 Segre types: type O, which is Einstein; type D$_\textrm{t}$, corresponding to $\mathbb{R} \times S^2$ or $\mathbb{R} \times H^2$, or D$_\textrm{s}$, corresponding to $\textrm{AdS}_2 \times \mathbb{R}$ or $\textrm{dS}_2 \times \mathbb{R}$; type N, which is the symmetric conformally flat Walker metric.

It remains to consider Segre type N spacetimes that solve TMG or are conformally flat.  In the TMG case, it was shown that such as spacetime must be Kundt \cite{Ahmedov:2010uk}, i.e.\ there is an expansion-free null geodesic congruence, which corresponds to the null decomposition of the type N traceless Ricci tensor.  In fact, a more general result is that any algebraically special TMG solution must be Kundt \cite{Nurowski:2015hwa}.  The derivation in \cite{Ahmedov:2010uk} shows that the Kundt condition also holds in the conformally flat case.


\subsection{Null warped AdS}


Kundt solutions of TMG were studied in detail in \cite{Chow:2009vt} (reused in Chapter 19 of \cite{Garcia-Diaz:2017cpv}).  We can check all of the possible type N solutions, to see which have a cyclic-parallel Ricci tensor.  These type N metrics can all be written in the form
\be
\df s^2 = \df \rho^2 + 2 \, \df u \, \df v + 	[v^2 f_2 (\rho) + f_0 (u, \rho)] \, \df u^2 + 2 W_1 (\rho) \, \df u \, \df \rho ,
\ee
where $W_1' = \tf{1}{2} W_1^2 + 2 \Lambda$, $f_2 = \tf{1}{4} W_1^2 + \Lambda$, and $f_0$ satisfies a linear third-order ordinary differential equation in $\rho$.  Note that, to match \cite{Chow:2009vt}, we reinstate dimensionful constants in this discussion.  We take $\Lambda = - m^2 < 0$ since the case $\Lambda = 0$ can be obtained as a limit and the case $\Lambda > 0$ can be obtained through analytic continuation.  However, the cases $\Lambda \geq 0$ do not lead to any metrics with 4 Killing vectors.  There are three possibilities for $W_1$: $W_1 = - 2 m$, $W_1 = - 2 m \coth (m \rho)$, and $W_1 = - 2 m \tanh ( m \rho)$, since a trivial constant of integration is removed by translation of $\rho$.  Within each of these cases for $W_1$, we have to consider separately the subcases $\mu \neq \pm m$, $\mu = m$ and $\mu = - m$.

Checking each of these cases shows that the only type N solutions with a cyclic-parallel Ricci tensor are within the class $W_1 = - 2 m$, $\mu \neq \pm m$.  Imposing the cyclic-parallel condition on all of the other cases implies the vanishing of one of the component functions of $f_0$, namely $f_{0 1} = 0$, and so these other cases give only pure AdS.  In contrast, for the $W_1 = - 2 m$, $\mu \neq \pm m$ case, for which the metric takes the form
\be
\df s^2 = \df \rho^2 + 2 \, \df u \, \df v - 4 m v \, \df u \, \df \rho + \expe{(m - \mu) \rho} f_{0 1} (u) \, \df u^2 ,
\ee
we find that
\be
\nabla_{(a} R_{b c)} \, \df x^a \, \df x^b \, \df x^c = \tf{1}{2} (m^2 - \mu^2) \expe{(m - \mu) \rho} [f_{0 1}'(u) \, \df u - (3 m + \mu) f_{0 1} (u) \, \df \rho] \, \df u^2 .
\ee
Neglecting the pure AdS$_3$ case $f_{0 1} = 0$, the cyclic-parallel condition $\nabla_{(a} R_{b c)} = 0$ implies that $\mu = - 3 m$ and $f_{0 1}$ is a non-zero constant.  By constant rescalings of $u$ and $v$, we may then set $f_{0 1} = \pm 1$, giving
\be
\df s^2 = \df \rho^2 + 2 \, \df u \, \df v - 4 m v \, \df u \, \df \rho \pm \expe{4 m \rho} \, \df u^2 .
\ee
The coordinate transformation $v \rightarrow \expe{2 m \rho} v$ puts the metric in the form
\be
\df s^2 = \df \rho^2 + 2 \expe{2 m \rho} \, \df u \, \df v \pm \expe{4 m \rho} \, \df u^2 ,
\ee
which corresponds to null warped AdS \eq{nullwarpedAdS}.


\subsection{Conformally flat Walker metrics}


We can follow the similar reasoning for the conformally flat metrics.  A Kundt metric can be expressed in canonical coordinates as
\be
\df s^2 = \df \rho^2 + 2 \, \df u \, \df v + f(v, u, \rho) \, \df u^2 + 2 W(v, u, \rho) \, \df u \, \df \rho .
\ee
For a Walker metric, the function $W$ can be taken to be independent of $v$, but we are not making that assumption here.  The Cotton tensor components $C_{v v}$ and $C_{v \rho}$ imply that
\begin{align}
W(v, u, \rho) & = v^2 W_2 (u, \rho) + v W_1 (u, \rho) + W_0 (u, \rho) , \nnr
f(v, u, \rho) & = v^2 f_2 (u, \rho) + v f_1 (u, \rho) + f_0 (u, \rho) .
\end{align}
The type N condition implies that $S{^a}{_b} S{^b}{_c} = 0$, from which one easily sees that $W_2 = 0$ and $f_2 = \tf{1}{2} \partial_\rho W_1$.

At this point, we note coordinate transformations that leave the general form of the Kundt metric unchanged:
\begin{align}
v & = \fr{\widetilde{v}}{\dot{u} (\widetilde{u})} + F(\widetilde{u}, \widetilde{\rho}), & u & = u(\widetilde{u}) , & \widetilde{\rho} = \widetilde{\rho} + G(\widetilde{u}) ,
\label{transform}
\end{align}
where $\dot{u} = \df u / \df \widetilde{u}$, and $(\widetilde{v}, \widetilde{u}, \widetilde{\rho})$ are the new coordinates.  The form of $W_1$ and $f_2$ is unmodified by these transformations, but $W_0$, $f_1$ and $f_0$ change.

If $W_1$ is independent of $\rho$, $W_1 = W_1(u)$, then the type N condition and the Cotton tensor component $C_{u \rho}$ imply that $W_1$ is a constant and $f_1$ is independent of $\rho$.  The function $F$ in the coordinate transformation \eq{transform} can be used to set $W_0 = 0$.  The Cotton tensor component $C_{u u}$ then implies that $f_0 (u, \rho) = f_{0 0} (u) + \expe{- W_1 \rho/2} f_{0 1} (u) + \expe{- W_1 \rho} f_{0 2} (u)$, at which point we now have a type N conformally flat metric.  However, imposing the symmetric condition $\nabla_a R_{b c} = 0$ or the condition $\nabla_a R_{b c} \, \nabla_d R_{e f} + 4 c^{-1} S_{a b} S_{c d} S_{e f} = 0$ implies that $W_1 f_{0 1} = 0$, which gives only an Einstein space.

If instead $W_1$ depends on $\rho$, then the function $G$ in the coordinate transformation \eq{transform} can be used to set $W_1 = W_1 (\rho)$, and then the function $F$ can be used to set $W_0 = 0$.  The type N condition implies that $f_1$ is independent of $\rho$, and then implies that $W_1'' - W_1 W_1' = 0$.  Removing a redundant integration constant, we obtain $W_1 = 2 k \, \tan (k \rho)$ for some constant $k$, related to the Ricci scalar as $R = 6 k^2$.

If $k \neq 0$, the Cotton tensor component $C_{u u}$ then implies that $f_0 (u, \rho) = \cos (k \rho) [f_{0 0} (u) + \sin (k \rho) f_{0 1} (u) + \cos (k \rho) f_{0 2} (u)]$, at which point we now have a type N conformally flat metric.  Again, imposing the symmetric condition $\nabla_a R_{b c} = 0$ or the condition $\nabla_a R_{b c} \, \nabla_d R_{e f} + 4 c^{-1} S_{a b} S_{c d} S_{e f} = 0$ gives only an Einstein space.

If instead we take $k = 0$, the Cotton tensor component $C_{u u}$ then implies that $f_0 (u, \rho) = f_{0 0} (u) + \rho f_{0 1} (u) + \rho^2 f_{0 2} (u)$, at which point we now have a type N conformally flat metric.  A reparameterization of $u$ in the coordinate transformation \eq{transform} can be used to set $f_1 = 0$.  A further coordinate \eq{transform} with $F(\widetilde{u}, \widetilde{\rho}) = F_2 (\widetilde{u}) - \df G/\df \widetilde{u}$ has two functions that can be used to set $f_{0 0} = 0$ and $f_{0 1} = 0$, so we arrive at
\be
\df s^2 = \df \rho^2 + 2 \, \df u \, \df v + \rho^2 f_{0 2} (u) \, \df u^2 .
\ee
We find that
\be
\nabla_a R_{b c} \, \df x^a \, \df x^b \, \df x^c = - 3 f_{0 2}' (u) \, \df u^3 .
\ee
If we demand that the spacetime is symmetric, then $f_{0 2}$ is a constant, which can be scaled to $\pm 1$, giving the symmetric Walker metric \eq{symmetricWalker}.  If we instead demand that $\nabla_a R_{b c} \, \nabla_d R_{e f} + 4 c^{-1} R_{a b} R_{c d} R_{e f} = 0$ for some $c \neq 0$, then $f_{0 2}'{^2} - 4 c^{-1} f_{0 2}^3 = 0$.  Removing a redundant integration constant gives $f_{0 2} = c/u^2$, so we obtain the non-symmetric Walker metric \eq{nonsymmetricWalker}.


\section{Comparison with literature}


The algebraic conditions on the curvature provide a simple way to check the equivalence of metrics, but sometimes one needs to be more explicit in demonstrating the equivalence.  To do this in practice, it can be helpful to find the eigenvectors of $S{^a}{_b}$ in the coordinate systems provided and to write the metric in a way that is adapted to the eigenvectors.  Furthermore, one can find the Killing vectors to match a canonical form of the isometry algebra.

The classification of 3-dimensional space(times) with 4 Killing vectors was first obtained by Kruchkovich \cite{kruchk}, where 8 metrics are presented at the end.  There is slight amendment of certain explicit metrics in \cite{kruchk2}, at the end of page 133.  We show here how the Kruchkovich metrics match the canonical metric forms provided above.

\paragraph{Metrics 1 and 2:} Metrics 1 and 2 (and metric 3 with $q \neq 0$ in \cite{kruchk2}) correspond to the non-symmetric Walker metric, and explicit coordinate transformations to a canonical form are given in \cite{Nozawa:2019dwu}.

\paragraph{Metric 3:} Metric 3, written in the form of \cite{kruchk2}, where it is metric 3 with $q = 0$, and making an overall sign change here, is
\be
\df s^2 = - 2 \, \df x_1 \, \df x_2 + \cos^2 x_1 \, \df x_3^2 = 2 \, \df x_1 \, (- \df x_2 + \mathcal{A}) + \df s_2^2 ,
\ee
where
\begin{align}
\mathcal{A} & = - \tf{1}{2} x_3^2 \, \df x_1 + x_3 \, \sin x_1 \, \cos x_1 \, \df x_3 , \nnr
\df s_2^2 & = x_3^2 \, \df x_1^2 - 2 x_3 \, \sin x_1 \, \cos x_1 \, \df x_1 \, \df x_3 + \cos^2 x_1 \, \df x_3^2 .
\end{align}
This form of the metric is derived by solving for the Killing vectors, identifying those that correspond to the canonical form of the isometry algebra, and comparing with the negative-sign symmetric Walker metric \eq{symmetricWalker}.  $\df s_2^2$ is flat $\mathbb{R}^2$ written in unusual coordinates, as verified by checking that its curvature vanishes, with volume-form
\be
\epsilon_2 = x_3 \, \cos^2 x_1 \, \df x_1 \wedge \df x_3 = \tf{1}{2} \df \mathcal{A} .
\ee
We can solve for the Killing vectors of $\mathbb{R}^2$ in this coordinate system, and identify a canonical pair, to derive Cartesian coordinates $x = x_3 \, \sin x_1 \, \cos x_1$ and $y = x_3 \, \cos^2 x_1$.

\paragraph{Metric 4:} Metric 4 is spacelike-squashed de Sitter.  The choices of $e_1$ and $e_3$ that lead to Lorentzian signature metrics are $(e_1, e_3) = (1, -1), (-1, 1), (-1, -1)$.  The corresponding metrics are respectively (with an overall sign change for the $(e_1, e_3) = (-1, -1)$ case):
\begin{align}
\label{Kruch4a}
\df s^2 & = - \df x_3^2 + \df x_1^2 + (\df x_2 + x_1 \, \df x_3)^2 , \\
\label{Kruch4b}
\df s^2 & = - \df x_1^2 + \df x_3^2 + (\df x_2 + x_1 \, \df x_3)^2 , \\
\label{Kruch4c}
\df s^2 & = - (\df x_2 + x_1 \, \df x_3)^2 + \df x_1^2 + \df x_3^2 .
\end{align}
\eq{Kruch4a} and \eq{Kruch4b} are spacelike-squashed flat space, presented as a fibration over a Minkowski$_2$ base spacetime.  \eq{Kruch4c} is timelike-squashed flat space, presented as a fibration over an $\mathbb{R}^2$ base space.

\paragraph{Metric 5:} Metric 5 contains various cases through choices of signs.  After absorbing some sign choices through the signs of $x_2$ and $x_3$, and overall sign changes of the metric, it gives
\begin{align}
\label{Krucha}
\df s^2 & = - (\df x_3 + n \expe{x_1} \, \df x_2)^2 + \df x_1^2 + (n^2 + 1) \expe{2 x_1} \, \df x_2^2 , \\
\label{Kruchb}
\df s^2 & = - (\df x_3 + n \expe{x_1} \, \df x_2)^2 + \df x_1^2 + (n^2 - 1) \expe{2 x_1} \, \df x_2^2 , \\
\label{Kruchc}
\df s^2 & = - (n^2 + 1) \expe{2 x_1} \, \df x_2^2 + \df x_1^2 + (\df x_3 + n \expe{x_1} \, \df x_2)^2 , \\
\label{Kruchd}
\df s^2 & = - (n^2 - 1) \expe{2 x_1} \, \df x_2^2 + \df x_1^2 + (\df x_3 + n \expe{x_1} \, \df x_2)^2 , \\
\label{Kruche}
\df s^2 & = - \df x_1^2 + (1 - n^2) \expe{2 x_1} \, \df x_2^2 + (\df x_3 + n \expe{x_1} \, \df x_2)^2 ,
\end{align}
where $n \geq 0$ in all cases, $n > 1$ in \eq{Kruchb} and \eq{Kruchd}, and $n < 1$ in \eq{Kruche}.  These are recognizable as timelike- and spacelike-squashed (A)dS, expressed as $\textrm{U}(1)$ bundles over a two-dimensional constant curvature metric.  \eq{Krucha} and \eq{Kruchb} are timelike-squashed AdS written, presented as fibrations over an $H^2$ base space.  \eq{Kruchc} and \eq{Kruchd} are spacelike-squashed AdS, presented as fibrations over an $\textrm{AdS}_2$ base spacetime.  \eq{Kruche} is spacelike-squashed dS, presented as a fibration over a $\textrm{dS}_2$ base spacetime.

A rescaling of $x_2$ puts \eq{Krucha}, \eq{Kruchb}, \eq{Kruchc} and \eq{Kruchd} into the forms
\begin{align}
\df s^2 & = - (\df x_3 + \nu \expe{x_1} \, \df x_2)^2 + \df x_1^2 + \expe{2 x_1} \, \df x_2^2 , \\
\df s^2 & = - \expe{2 x_1} \, \df x_2^2 + \df x_1^2 + (\df x_3 + \nu \expe{x_1} \, \df x_2)^2 ,
\end{align}
where $\nu \geq 0$ and the timelike- and spacelike-squashings are separated by $\nu = 1$, which is a pure AdS metric.

\paragraph{Metric 6:} For Lorentzian signature metrics, we need to take $e_1 = 1$ and $m < 0$ or $e_1 = -1$ and $m > 0$.  However, these are equivalent when the overall sign of the latter metric is changed, and lead to
\be
\df s^2 = - m (\df x_3 - \sin x_1 \, \df x_2)^2 + \df x_1^2 + \cos^2 x_1 \, \df x_2^2 ,
\ee
with $m > 0$.  This is timelike-squashed dS, presented as a fibration over an $S^2$ base space.

\paragraph{Metric 7:} This gives $\mathbb{R} \times S^2$.  If $x_3$ is analytically continued, then we obtain $\textrm{dS}_2 \times \mathbb{R}$.

\paragraph{Metric 8:} This gives $\mathbb{R} \times H^2$ and $\textrm{AdS}_2 \times \mathbb{R}$.


\section{Conclusion}


We have summarized the classification of 3-dimensional Lorentzian metrics that locally admit 4 independent Killing vectors.  They have been characterized in terms of algebraic conditions on the curvature and its derivative.  These conditions are of practical use as an initial check on a test metric, without the need to compute eigenvectors and explicitly decompose the metric into a triad.  It may be possible to generalize the types of algebraic conditions on curvature and its derivatives provided here to metrics with slightly fewer Killing vectors, or to metrics in higher dimensions.

\end{document}